\documentclass[aps,prl,amsmath,amssymb,reprint,superscriptaddress,]{revtex4-1}

    \usepackage{physics}
    \usepackage{graphicx}
    \usepackage{grffile}
    \usepackage{float}
    \usepackage{color,soul}
    \usepackage{hyperref}
    \usepackage{xspace}
    \usepackage{makecell}
    \usepackage{comment}
    \usepackage[version=4]{mhchem}
    \usepackage{xstring}
    \usepackage{xcolor}
    \usepackage[flushleft]{threeparttable}

    \usepackage{makeidx}



    \newcommand{\kel}{\ensuremath{\kappa_{\textrm{el}}}}
    \newcommand{\PLZ}{\ensuremath{P_{\textrm{LZ}}}}
    \newcommand{\Hab}{\ensuremath{H_{\textrm{ab}}}}
    \newcommand{\veff}{\ensuremath{\nu_{\textrm{eff}}}}

\begin{document}

\title[]{Combining Landau Zener Theory and Kinetic Monte Carlo Sampling for 
Small Polaron Mobility of Doped \ce{BiVO4}}
\author{Feng Wu}
\affiliation{The Department of Chemistry and Biochemistry, University of California, Santa Cruz, 95064 CA, United States}
\author{Yuan Ping}
\email{yuanping@ucsc.edu}
\affiliation{The Department of Chemistry and Biochemistry, University of California, Santa Cruz, 95064 CA, United States}
\date{Apr 15, 2018} 

\begin{abstract}
\end{abstract}

\begin{abstract}
    Transition metal oxides such as \ce{BiVO4} are promising materials as photoelectrodes in solar-to-fuel conversion applications. However, their performance is limited by the low carrier mobility (especially electron mobility) due to the formation of small polarons. Recent experimental studies show improved carrier mobility and conductivity by atomic dopings; however the underlying mechanism is not understood. A fundamental atomistic-level understanding of the effects on small polaron transport  is critical to future materials design with high conductivity. We studied the small polaron hopping mobility in pristine and doped \ce{BiVO4} by combining Landau-Zener theory and kinetic Monte Carlo (kMC) simulation fully from first-principles, and investigated the effect of dopant-polaron interactions on the mobility. We found polarons are spontaneously formed at V in both pristine and Mo/W doped \ce{BiVO4}, which can only be described correctly by density function theory (DFT) with the Hubbard correction (DFT+U) or hybrid exchange-correlation functional but not local or semi-local functionals. 
    We found DFT+U and dielectric dependant hybrid functional (DDH) give similar electron hopping barriers, which are also similar between the room temperature monoclinic phase and the tetragonal phase.  
    The calculated electron mobility agrees well with experimental values, which is around $10^{-4}$ cm$^2$V$^{-1}$s$^{-1}$. We found the electron polaron transport in \ce{BiVO4} is neither fully adiabatic nor nonadiabatic, and the first and second nearest neighbor hoppings have significantly different electronic couplings between two hopping centers that lead to different adiabaticity and prefactors in the charge transfer rate, although they have similar hopping barriers. Without considering the detailed adiabaticity through Landau-Zener theory, one may get qualitatively wrong carrier mobility. We further computed polaron mobility in the presence of different dopants and showed that Cr substitution of V is an electron trap while Mo and W are "repulsive" centers, mainly due to the minimization of local lattice expansion by dopants and electron polarons. The dopants with "repulsive" interactions to polarons are promising for mobility improvement due to larger wavefunction overlap and delocalization of locally concentrated polarons.
\end{abstract}

\maketitle

    \section{Introduction}
    
    Transition metal oxides (TMO) such as \ce{BiVO4}, \ce{Fe2O3}, \ce{CuO} are promising candidates as photoelectrode materials in energy conversion applications, such as photoelectrochemical cells\cite{Walter2011,Sivula2016,Mi2012,Hill2013,Kim2015,Pham2017,Ping2012,Ping2013,Ping2013a,Ping2014,Smart2017}, due to their high stability under electrochemical conditions compared to III-V semiconductors and desired optical properties for visible light absorption. However, in general, these oxides have extremely low intrinsic carrier mobility (\textit{e.g.} on the order of 0.01$ \textrm{cm}^2\textrm{V}^{-1}\textrm{s}^{-1}$ hole mobility for \ce{BiVO4}\cite{Rettie2013} compared to 1350 $\textrm{cm}^2\textrm{V}^{-1}\textrm{s}^{-1}$ for silicon\cite{Li1977}), which fundamentally limits their efficiency from the theoretical value, and constitutes the main bottleneck of these materials for practical applications. The extremely low carrier mobility is characterized by the thermally activated hopping conduction\cite{Kim2015,Smart2017}, instead of band conduction in III-V semiconductors. 
            
\begin{figure}
        \centering
        \includegraphics[width=0.5\linewidth]{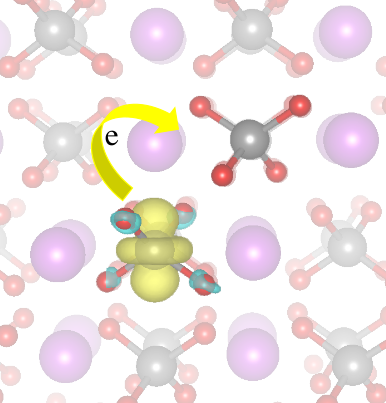}
        \caption{Small polaron hopping in \ce{BiVO4}. The yellow isosurface is the spin density of the polaron. The isosurface is 0.0045 e/Bohr$^3$. Silver ball:V atoms, red ball:O, purple ball: Bi.} 
        \label{fig:BiVO4-hopping-spin}
    \end{figure}

    The carriers in the hopping conduction of TMOs are called "small polarons", which are quasiparticles of electron plus local lattice distortion as a whole. Its formation is due to the extremely strong electron-phonon interactions, whereby the electrons or holes are trapped by local lattice distortions, and they hop from one lattice site to another. A spin density plot of an electron small polaron in pristine \ce{BiVO4} is shown in Fig. \ref{fig:BiVO4-hopping-spin}. Experimentally, a distinct signature of polaron hopping conduction is that with increasing temperature the carrier mobility increases exponentially, while in band conduction it decreases. A linear dependence between the logarithmic conductivity and temperature is often observed experimentally in polaronic materials, where the slope of  linear dependence is the hopping activation energy.
    
    Interestingly, it has been observed that certain dopants in TMOs can improve their carrier mobility by lowering the polaron hopping barriers (activation energies). For example, in the case of N-doped \ce{BiVO4} with excessive oxygen vacancies, both the carrier concentration and mobility can be enhanced~\cite{Kim2015}. In particular, formation of N-V bonds decreased the static dielectric constant and lowered the hopping barriers of polaron transport. 
    Similarly, the hopping barrier can be significantly lowered by Li doping in CuO\cite{Smart2018,Zheng2000,Zheng2004}, \textit{i.e.} the hopping barriers in CuO decreased an order of magnitude after 16\% Li doping, due to a combined effect of lowering electron-phonon interaction and magnetic coupling after doping. Note that the carrier conductivities depend on the hopping barrier exponentially $\sigma\propto\exp (-E_a/k_BT)$ \textit{e.g.} decreasing the hopping barrier by 25 meV can lead to three times improvement on carrier mobility. Furthermore, recent experimental work shows Mo/W doping can increase the photocurrent of \ce{BiVO4} \cite{Zhang2018,Abdi2013,Abdi2013a,Abdi2017,Rettie2013,Rettie2015,Rettie2016,Ziwritsch2016}. The photocurrent is proportional to the product of carrier concentration and carrier mobility (the optical absorption could also affect photocurrents but it has been shown unchanged after W and Mo doping~\cite{Luo2013}). The carrier concentration has been shown to increase due to the shallow nature of Mo/W dopants in BiVO$_4$\cite{Luo2013,Rohloff2017}; however, whether the electron mobility of BiVO$_4$ increases after Mo/W doping is undetermined. Some studies showed a lowered mobility in Mo/W doped \ce{BiVO4}\cite{Abdi2013,Ziwritsch2016} while others suggested a lowered activation energy of conduction and improved carrier mobility\cite{Zhang2018}. Overall, these studies suggest the possibility of overcoming slow electronic conduction in these TMOs by appropriate atomic doping. 
    
    However, to date, although there are several important related discussions\cite{Hoang2018}, there is still an incomplete understanding of the doping effect on small polaron formation and mobility in TMOs, both theoretically and experimentally. Further improvements of conductivities in these TMOs require rationale design of effective dopants, which need reliable \emph{ab-initio} tools to make predictions for small polaron mobility. 
    
    Previous computational methods of small polaron mobility have relied on applying the Marcus theory in the context of polaronic systems or Emin-Holstein-Sustin-Mott theory (EHAM)\cite{EHAM1,EHAM2,EHAM3}. Despite the significant progress that has been made in the calculations of small polaron mobility\cite{Adelstein2014,Deskins2007,Alidoust2015,Liao2011,Liu2018,Oberhofer2012,Oberhofer2017}, several major limitations still remain: a) most studies for solid systems computed the hopping rates at the adiabatic limit\cite{Adelstein2014,Deskins2007,Kweon2015,Plata2013}, which may not always be valid, especially for magnetic TMOs\cite{Smart2018}; b) the prefactor for polaron hopping rates was rarely computed~\cite{Wang2016,Adelstein2014}, and an estimated value was often used without detailed justification\cite{Kweon2013,Liu2018}; c) a simple analytic formula based on the assumption of isotropic hopping in solids with the same hopping rates for each hop was mostly used, 
    which is fundamentally not applicable to doped solids or systems with low symmetry.
    
    In this paper, we will \emph{first} introduce our recent development on first-principles calculations of small polaron hopping mobility by combining Landau-Zener theory including both adiabatic and non-adiabatic electron transfer with a kinetic Monte Carlo (kMC) sampling or specifically random walk sampling (RWS) method; \emph{next} we will discuss how we apply this method to compute hopping mobility in pristine \ce{BiVO4} and discuss its dependence on the level of theory and the hopping range; \emph{at the end}, we will show how the dopants affect the polaron energies and hopping mobility through our kMC sampling, and suggest the design principles of "good dopants" that can boost small polaron mobility of TMOs.
    
    \section{Theory and Computational Methods on Small Polaron Hopping Mobility}
    The theory for small polaron rates fundamentally relies on the fact that there is a non-zero barrier for electron/hole hopping from the initial site to the final site, where the “site” is defined by the charge localization volume in solids with a few angstrom radius. The small polaron hopping transport is analogous to the charge transfer in a molecular crystal where the charge is highly localized on a few atoms or one molecule at each hop. Our discussion of the theoretical methodology will start with the definition of carrier mobility, then its relationship to the diffusion coefficient $(D)$ by the Einstein-Smoluchowski equation at the weak electric field limit and $D$'s relation to hopping transfer rates $k_{ET}$ by kMC samplings, and afterward computing $k_{ET}$ by the generalized Landau-Zener theory, where first-principles approaches to compute each part in the formulation will be introduced.
    
    The carrier mobility is defined as the velocity response of a charge carrier to an external electric field: 
    \begin{equation}
        \mu_{ij}=\frac{\expval{v}_i}{E_j}
        \label{eqmu}
    \end{equation}
    
    where $\expval{v}_i$ denotes the $i$-th component of the time-averaged velocity $\expval{v}$ of the carrier and $E_j$ is a component of the electric field vector $E$. In the regime of weak electric field (the regime we usually study), the carrier mobility can be expressed by the Einstein-Smoluchowski (ES) equation:
    \begin{equation}
    \mu_{ij}^{ES}=\frac{D_{ij}q}{k_B T}
     \label{eqES}
    \end{equation}
    where $D_{ij}$ is the diffusion coefficient tensor and q is the carrier charge. The diffusion coefficient tensor $D_{ij}$ follows a generalization of Fick’s law to velocity $v$ at time $t$. $D_{ij}$ is related to the electron transfer rate $k_\textrm{ET}$ at each hopping process ($D\propto k_{\textrm{ET}}$). For isotropic systems, a geometric factor could be used to relate $D$ and $k_\textrm{ET}$. For non-isotropic systems this needs to be sampled statistically which we will discuss later. \cite{Adelstein2014,Deskins2007,Kweon2015}
    
    With the harmonic approximation, the electron transfer rate in the Landau-Zener (LZ) theory\cite{Landau1932,Landau1932a,Zener1932} extended with nuclear quantum effects is:\cite{Brunschwig1980,Newton1991}
    \begin{equation}
    k_{\textrm{ET}}=\kel\veff\Gamma\exp(-E_a/k_BT)
    \label{eq:lz-rate}
    \end{equation}
    where $\kel$ and $\Gamma$ are the thermally averaged electronic transmission coefficient and nuclear tunneling factor respectively (taking into account the quantum effects of nuclear degree of freedom; but we will approximate $\Gamma\approx$1  in this study, since it's only important for low temperature or light elements). $\veff$ is the effective frequency along the reaction coordinate of electron transfer, $E_a$ is the hopping activation energy, regardless of adiabatic or non-adiabatic processes. 
    
\begin{figure}
        \centering
        \includegraphics[width=0.8\linewidth]{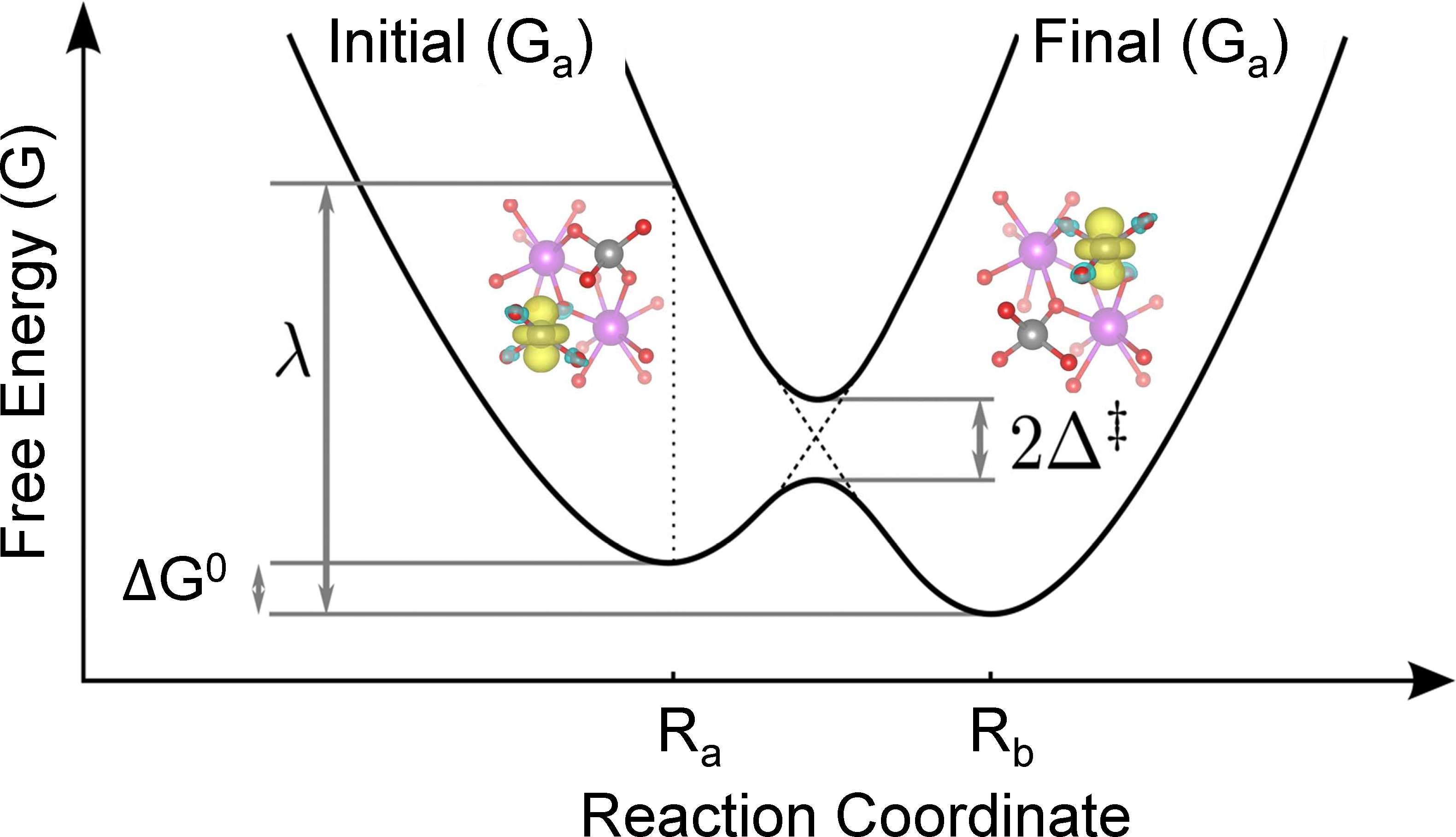}
        \caption{Electron hopping diagram along one dimensional configuration coordinate. The polaron spin densities with the yellow isosurface for the initial a and final b structures are shown (only local structures of the solid are shown here).} 
        \label{fig:lz-hopping}
    \end{figure}

    The electronic transmission coefficient $\kel$ represents the probability of electron transfer when the nuclear configuration approaches the intersection region where the transfer may happen.\cite{Newton1991} $\kel$ that corresponds to the situation when the crossing point is between the two potential wells follows
    \begin{equation}
        \kel=2\PLZ/(1+\PLZ)
    \end{equation}
    where $\PLZ$ is the Landau-Zener transition probability for a single potential energy surface crossing event (see Fig. \ref{fig:lz-hopping}), 
    \begin{equation}
        \PLZ=1-\exp(-2\pi\gamma)
    \end{equation}
    And $\gamma$ is the adiabaticity parameter defined as 
    \begin{equation}
        \label{eq:lz-2pigamma}
        2\pi\gamma=\frac{\pi^{\frac{3}{2}}\left|\Hab \right|^2}{h\veff\sqrt{\lambda k_B T}}
    \end{equation}
    where $h$ is Planck constant and $\Hab=\mel{\Psi_a}{H}{\Psi_b}_{TS}$ is the Hamiltonian transition matrix element or electronic coupling between initial $a$ and final $b$ electronic states at the transition state equilibrium geometry (TS), and $\lambda$ is the reorganization energy as shown in Fig. \ref{fig:lz-hopping}. The deviation of $\kel$ or  $\PLZ$ from unity is generally interpreted as a non-adiabatic behavior. Note that when $\PLZ(\kel)\rightarrow 1$, the Landau-Zener theory is reduced to the classical transition state theory; and when $\PLZ\rightarrow 0$, it is reduced to the Marcus theory~\cite{Marcus1993}.
    
    In principles, once one obtained $E_a$, $\veff$, $\Hab$, $\lambda$ (if we assume nuclear tunneling factor $\Gamma=1$), the small polaron hopping rates can be computed based on Eq.\ref{eq:lz-rate}. In practice, these calculations have rarely been carried out for extended solid state systems up to now. Most calculations have been performed with finite cluster models~\cite{Liao2011,Plata2013,Deskins2007,Liu2018}, or hopping transfer rates have been obtained at either adiabatic or nonadiabatic hopping limit~\cite{Kweon2013,Liu2018}, or $\Hab$ has been estimated from the energy difference between bonding and anti-bonding polaron states computed by DFT~\cite{Adelstein2014,Wang2016}, which may suffer from the DFT band gap problems.
    Next, we will introduce how we compute each part in Eq.\ref{eq:lz-rate} to Eq.\ref{eq:lz-2pigamma}, and then how we obtain the charge transfer rates in Eq.\ref{eq:lz-rate} and carrier mobility in solids. 
    
    \textbf{Activation Energy $E_a$} - it can be obtained through several theoretical methods depending on adiabatic or non-adiabatic processes. A general form independent on the adiabaticity is $E_a=\Delta E^{\ddagger}-\Delta^{\ddagger}$ (Eq.~\ref{eq:hopping-energy-correction}), where $\Delta E^{\ddagger}$ is the activation energy on the diabatic potential energy surface and $\Delta^{\ddagger}$ is a correction factor relating $\Delta E^{\ddagger}$ to the activation energy on the adiabatic potential energy surface (including the electronic coupling between initial and final states, as shown in Fig.\ref{fig:lz-hopping})\cite{Oberhofer2012}. The reaction coordinate \textbf{R} in Fig. \ref{fig:lz-hopping} represents a collective variable describing relaxation of the surrounding medium to changes in a local charge state. Previous studies have shown that this one-dimensional configuration coordinate can successfully describe the small polaron hopping and hopping activation energies of TMOs\cite{Adelstein2014,Kweon2015}. $\Delta E^{\ddagger}$ and $\Delta^{\ddagger}$ can be obtained by:
    \begin{align}
        \Delta E^{\ddagger} &= \frac{(\lambda+\Delta G^0)^2}{4\lambda} \label{eq:hopping-energy-diabatic}\\
        \Delta^{\ddagger} &= |\Hab|+\frac{\lambda+G_0}{2}-\sqrt{\frac{(\lambda+\Delta G^0)^2}{4}+|\Hab|^2}\\
        E_a&=\Delta E^{\ddagger}-\Delta^{\ddagger}
        \label{eq:hopping-energy-correction}
    \end{align}
    where $\Delta G^0$ is the energy difference between the minima of the two diabatic potential energy surfaces $a$ and $b$ (which can also be called ``driving force" for the electron transfer), and $\lambda$ is the reorganization energy as shown in Fig. \ref{fig:lz-hopping}. 
    
    In this paper we compared the barriers $E_a$ obtained with several approaches: Climbing Image-Nudged Elastic Band (NEB) approach~\cite{VTSTNEB1} through which the barrier is defined as the difference between the initial state and the transition state (saddle point) with both electronic and ionic relaxation, the commonly used linear interpolation (LERP) of configurations between the initial and final polaron states with linearly interpolated atomic positions and only electronic relaxation\cite{Maxisch2006} and the barrier is defined between the highest energy along the pathway and the energy of initial state, and Constrained Density Functional Theory (CDFT) method to obtain barriers based on Eq. \ref{eq:hopping-energy-diabatic} and \ref{eq:hopping-energy-correction}, with a new implementation for solids~\cite{CDFT}. This method has been recently applied to calculating polaron hopping barriers of metal oxides.\cite{hosung2018}
    
    \textbf{Effective Frequencies $\veff$} - we obtained it through  transition state theory with harmonic approximations, when a transition state can be well-defined. We note that for cases without a well-defined transition state (which means a non-adiabatic charge transfer process), the Marcus theory formula is used instead: $k=\frac{2\pi}{\hbar}\frac{1}{\sqrt{4\pi\lambda k_BT}}\left|\Hab \right|^2\Gamma \exp(\frac{-(\Delta G^0+\lambda)^2}{4\lambda k_B T})$, where an effective frequency is not necessary. The former case obtained through transition state theory assumes that a hopping process proceeds over a transition state, which is in thermodynamic equilibrium with its surroundings. The vibrational degrees of freedom at the transition state and the initial state determine the partition function. \emph{Ab initio} phonon calculations provide all vibrational terms, \textit{i.e.} the zero-point energy, temperature dependent part of the internal energy and vibrational entropy, taking into account the full coupling of the vibrational modes between the polarons and the host lattice. The effective frequency entering the rate equation Eq. \ref{eq:lz-rate} and Eq. \ref{eq:lz-2pigamma} is given by\cite{Wimmer2008,Adelstein2014}:
    \begin{align}
        \nu_{\textrm{eff}} &= \frac{k_BT}{h}
            \frac{Z_{\textrm{TS}}}{Z_{\textrm{GS}}} \nonumber\\
        &= \frac{k_BT}{h}\frac{\prod_i^{3N-6}\left[
            2\sinh\left(\frac{h\nu_i^{\textrm{GS}}}{2k_BT}\right)\right]}{
           \prod_i^{3N-7}
        \left[2\sinh\left(\frac{h\nu_i^{\textrm{TS}}}{2k_BT}\right)\right]} \label{eq:effective-freq-phonon}
    \end{align}
    
    where $Z_{TS}$ and $Z_{GS}$ are partition functions for the transition state and the ground state, respectively; $\nu_i$ are vibrational eigenmodes of the corresponding geometry. The details of geometry optimization and phonon calculations can be found in SI. 
    
    \textbf{Kinetic Monte Carlo simulation for $D$} - in order to accurately take into account of the anisotropic polaron hopping in pristine and doped systems, we implemented the kMC sampling to simulate the diffusion coefficients and hopping mobility in doped TMOs.
    The diffusion coefficient can be expressed as\cite{Yang2008,Oberhofer2017,Shuai2014}
    \begin{equation}
        \label{eq:diffusion-rw}
        D = \lim_{t\rightarrow \infty}\frac{\expval{L(t)^2}}{2Nt}
    \end{equation}
    where $N$ is the dimensionality of the kMC process, $\expval{L(t)^2}$ is the mean squared displacement (MSD) and $t$ is the time. The MSD is determined by the hopping rate $k_{\textrm{ET}}$ and the distance 
    between two lattice sites for each hop included in the kMC simulation. (Details of the algorithm and numerical tests can be found in SI.)
    Afterward we can obtain hopping mobility through Einstein-Smoluchowski(ES) equation in Eq.\ref{eqES}.
    The main advantage of the statistical sampling approach above over the analytic solution used in the past work is that it takes into account different hopping rates statistically and, most importantly, can also be applied to disordered and defective systems, which have significant value for practical applications. 
    The electronic structure and geometry relaxation calculations are performed in the open source plane wave code Quantum-ESPRESSO~\cite{QE2} by using norm-conserving pseudopotentials~\cite{ONCV2} with several exchange correlation functionals, as will be discussed later. More computational details can be found in SI.
    \section{Results and Discussions}
    
    \subsection{Small Polaron Hopping Conduction of Pristine \ce{BiVO4}}
    
    \subsubsection{Activation barriers $E_a$ with different theoretical methods}
    The most important quantity for small polaron hopping rates $k_{ET}$ and mobility is the hopping activation barrier ($E_a$) in Eq.\ref{eq:lz-rate}, due to its exponential relationship to $k_{ET}$. We will examine $E_a$ of \ce{BiVO4} with various computational methods in this section.
    The stable room temperature phase of \ce{BiVO4} is monoclinic, which has a very similar atomic structure with its high temperature tetragonal phase\cite{Sleight1979}. The tetragonal phase consists of \ce{VO4} and \ce{BiO8} polyhedra, with only one set of V-O bond length and two sets of Bi-O bond lengths. Each oxygen atom is three coordinated with one V and two Bi atoms. The monoclinic phase structure can be viewed as a slightly distorted tetragonal phase structure, and the V-O bond lengths are split into two groups at the two sides of V (bond length splitting, BLS). Consistent with the past work\cite{Park2011,Kweon2012}, we found at both  the DFT+U and PBE levels of theory, all V-O bond lengths become very close and the BLS at the monoclinic phase cannot be correctly described; instead, by increasing the exact exchange ratio above 10\%, the experimental monoclinic BLS can be reproduced\cite{Kweon2012}.
    The previous first-principles calculation of \ce{BiVO4} band structure shows that the effect of BLS (or the difference between tetragonal and monoclinic phase structures) is mainly important at the valence band maximum (VBM), but the conduction band minimum is weakly affected\cite{Kweon2012}, which indicates the BLS may have minimum effects on the electron conduction compared to the hole conduction, as discussed below. 
    
    To understand the difference of electron transport between tetragonal and monoclinic phases, we further investigated how the hopping barriers depend on the tetragonal and monoclinic structures, along with the comparison between different DFT functionals and theoretical methods for activation barriers (\textit{i.e.} NEB, LERP, CDFT), 
    as shown in Table \ref{table:compare-barrier-method} and summarized below. 
    Firstly, we found the hopping barriers increase with the fraction of Fock exchange $\alpha$ in hybrid functionals (253 meV at $\alpha$=0.1449; 357 meV at $\alpha$=0.25) based on Constrained DFT~\cite{CDFT}. This is a general physical effect independent of the specific system we study: increasing $\alpha$ in hybrid functionals will increase the electronic wavefunction localization and lower the electronic coupling between two hopping sites (lower $\Hab$), and therefore increase the hopping barriers. At the limit of $\alpha=0$ (at the PBE level), we cannot obtain a positive hopping barrier or localized small polaron state, due
    to the charge delocalization error in DFT semi-local  functionals. Therefore, 
    PBE does not describe the conduction of \ce{BiVO4} as an activated polaron hopping
    which is fundamentally contradictory to the experimental conductivity measurements, and should not be used to describe the electronic structure and carrier transport in \ce{BiVO4}. 
    
    Secondly, the hopping barriers are very similar between tetragonal and monoclinic phases by CDFT at the same level of theory, specifically dielectric dependent hybrid functional (DDH) where $\alpha$ depends on the inverse of high frequency dielectric constant $\epsilon_{\infty}$, with $\alpha=0.1449$ (computed $\epsilon_{\infty}=6.9$) for \ce{BiVO4}~\cite{Kim2015}. 
    
\begin{figure}
        \centering
        \begin{minipage}{.45\linewidth}
        \includegraphics[width=1.0\linewidth]{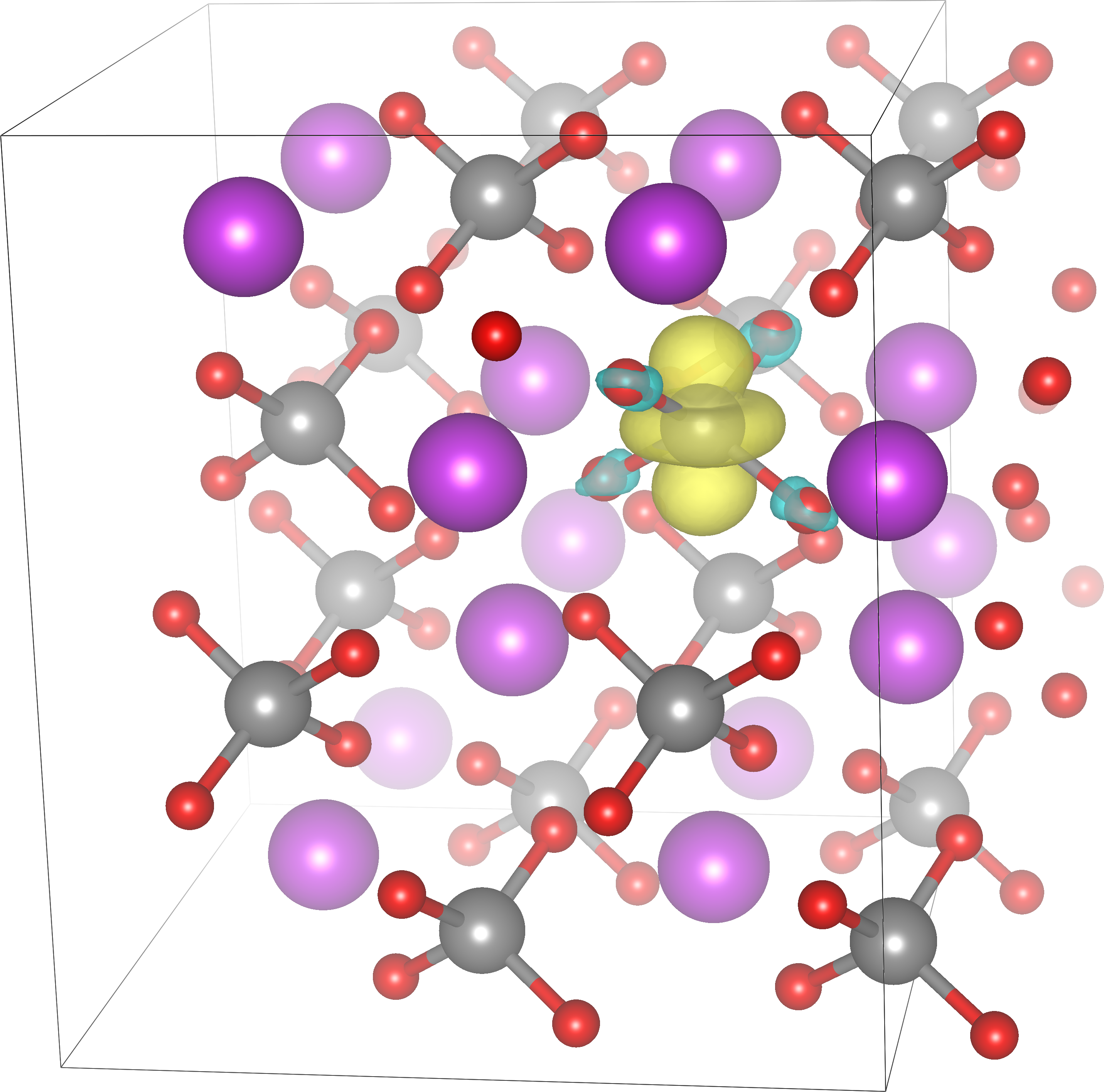}
        \end{minipage}
        \begin{minipage}{.45\linewidth}
        \includegraphics[width=1.0\linewidth]{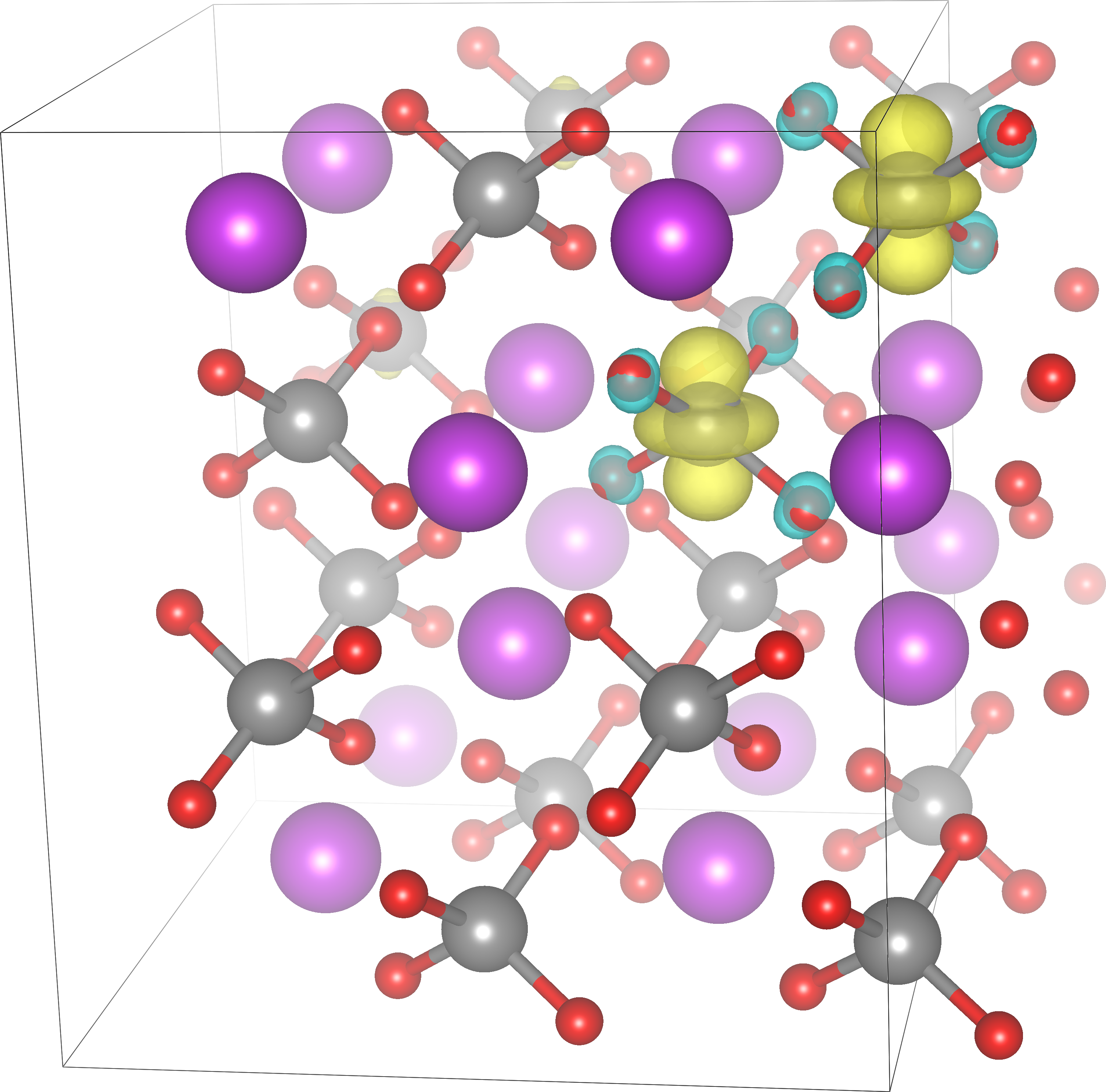}
        \end{minipage}
        \caption{Spin density plots of polaronic ground state (left) and hopping transition state (right). We showed that the spin density of the excess electron is well localized inside the \ce{VO4} tetrahedral at the ground state, and the transition state is simply a combination of two  localized half-electron on two sites. The isosurface is 0.0045 e/Bohr$^3$.}
        \label{fig:spindensity-polaron-gs-ts}
    \end{figure}

    Thirdly, DFT+U and DDH give similar barriers within 40 meV (computed with CDFT). As DDH generally provides reliable electronic structure and polaronic properties for bulk systems\cite{Brawand2017,Skone2014}, the similar results between DDH and DFT+U (V(U)=2.7 eV based on past work~\cite{Park2011,Kim2015}) show the reliability of DFT+U calculations for the hopping barriers of \ce{BiVO4}, which is also more computationally affordable. Therefore, we used DFT+U for barrier calculations with other methods as well, such as Climbing Image-Nudged Elastic Bands (NEB) and Linear Interpolation (LERP). Note that both NEB and LERP assume the adiabaticity of the charge transfer process; namely a well-defined transition state is necessary to define the barrier height. Indeed, we found a well-defined transition state of the nearest neighbor hopping in \ce{BiVO4}, where the spin density is distributed equally on two hopping sites (see Fig. \ref{fig:spindensity-polaron-gs-ts}), which proves the validity of NEB and LERP methods. Indeed, CDFT, NEB and LERP give similar barriers (217, 247, 257 meV respectively) for the monoclinic phase at DFT+U level of theory.
    Therefore, we mostly used the NEB method with DFT+U for the barrier calculations in the rest of this paper for a good balance between accuracy and computational
    cost.
    On the other hand, the parabola fitting which neglects the electron coupling between two diabatic states will significantly overestimate the barrier of this adiabatic process (546 meV by this work, and 460 meV by Ref~\citenum{Kweon2015}, strongly overestimated compared with 357 meV in CDFT with the same functional PBE0).
    
    \begin{table}
        \centering
        \caption{Polaron hopping activation barriers of nearest neighbor hopping in pristine tetragonal and monoclinic \ce{BiVO4}, computed with four different methods, including Constrained DFT (CDFT), Nudged Elastic Bands (NEB), Linear Interpolation (LERP) and Parabola Fitting (Parabola Fit), at DFT+U (U(V)=2.7 eV), dielectric dependent hybrid functional (DDH, $\alpha=0.1449$) and PBE0 ($\alpha=0.25$). Note that all the geometries are optimized at the corresponding level of theory; at DFT+U, the monoclinic phase does not have a bond length splitting (BLS), unlike at hybrid functionals.}
        \label{table:compare-barrier-method}
        \begin{tabular}{cccc}
          \hline
          Method & DFT method & Phase & Barrier (meV)\\
          \hline
          CDFT & Hybrid-DDH & Tetragonal & 253 \\
          CDFT & Hybrid-DDH & Monoclinic & 249 \\
          CDFT & Hybrid-PBE0 & Monoclinic & 357 \\
          CDFT & DFT+U & Monoclinic & 217 \\
          NEB & DFT+U & Monoclinic & 247 \\
          LERP & DFT+U & Monoclinic & 257 \\
          Parabola Fit & Hybrid-PBE0 & Monoclinic & 546 \\
          Parabola Fit & Hybrid-PBE0 & Monoclinic & 460\cite{Kweon2012} \\
          \hline
        \end{tabular}
    \end{table}
    
    \subsubsection{Effective frequencies $\veff$ and charge transfer rates $k_{\textrm{ET}}$}
    
    All parameters used in Landau-Zener theory for charge transfer rates $k_{\textrm{ET}}$ are computed and summarized in Table \ref{table:hopping-parameters}. 
    We examined both the first (1NN) and second nearest neighbor (2NN) hoppings as we found their hopping barriers are comparable (see Table \ref{table:hopping-parameters}); and a recent work\cite{Rettie2015} claimed a second
    nearest neighbor hopping may have significant contribution to the hopping mobility in \ce{BiVO4}.
    $\Hab$ and $\lambda$ are computed from CDFT in a supercell of 192 atoms with DFT+U. Due to the high computational cost, the effective frequencies were computed from $\Gamma$-point phonons of the ground state and transition state in a supercell of 96 atoms by Eq.\ref{eq:effective-freq-phonon}. Details of phonon frequency calculations and the effective frequency from  classical high-temperature limit are provided in SI. The effective frequency depends on the temperature, but in Table \ref{table:hopping-parameters} we show values for 300K only in order to be consistent with the temperature we simulate the polaron mobility later. 
    
    From Table. \ref{table:hopping-parameters} we can see the transfer probability $\PLZ$ is 0.6 for the first nearest neighbor (1NN) hopping, which is on the borderline of adiabatic hopping $\PLZ\rightarrow 1$ and nonadiabatic hopping $\PLZ\rightarrow 0$, and closer to the adiabatic one. Meanwhile, the second nearest (2NN) hopping has $\PLZ=0.1$, which is small and we could state that this process is closer to nonadiabatic.
    The small $P_{LZ}$ for the 1NN compared to the 2NN is because the adiabaticity parameter in Eq.6 is proportional to the square of electronic coupling $H_{ab}$, which is in turn proportional to the overlap of electronic wavefunctions between two hopping sites. For both 1NN and 2NN hoppings, the electron is localized on VO$_4$. However, the hopping distances are different (i.e. 3.9 Å for 1NN and 5.2 Å for 2NN), and the wavefunction overlap decreases exponentially with distances that lead to small Hab and low $P_{LZ}$ for 2NN. Interestingly, for the 2NN hopping, we still found a well-defined transition state and a barrier of 269 meV obtained by NEB, similar to the barrier obtained by CDFT (294 meV) where adiabaticity
    is not assumed in Eq.~\ref{eq:hopping-energy-correction}. Note the transition from adiabatic to non-adiabatic does not have a clear boundary, so the $\PLZ$ at which the transition state cannot be defined is undetermined.
    Neither 1NN nor 2NN $\PLZ$ is very close to 0 or 1, which means the polaron hopping in \ce{BiVO4} is neither completely adiabatic nor nonadiabatic, and demonstrates the importance of applying the Landau-Zener theory here instead of classical transition state theory (which is only valid for an adiabatic process) or Marcus theory (which is only applicable to a nonadiabatic process). 
    
    The main difference between the first and second nearest neighbor hopping is the electron coupling matrix $\Hab$, for which 1NN hopping is 4 times larger than that of 2NN hopping. This is a direct consequence of a longer hopping distance (\textit{i.e.} 5.0 \AA\, in 2NN compared to 3.9 \AA\, in 1NN):
    as the polaron localization length stays the same, the wavefunction overlap between two hopping sites is strongly reduced due to the exponential decay of wavefunctions, so does a strong reduction of $\Hab$. This results in the adiabaticity parameter $\gamma$ 16 times smaller in 2NN than 1NN due to the $|\Hab|^2$ term in Eq.\ref{eq:lz-2pigamma}. The difference of adiabaticity between different hoppings in the same system can also be found in other materials, like the intralayer hopping and interlayer hopping in \ce{FePO4}, which has a layered structure.\cite{Wang2016}
     
    \begin{table}
        \centering
        \caption{Key parameters computed fully from first-principles for the charge transfer rate at 300K of the first (1NN) and second nearest neighbor (2NN) hopping studied in this work: electron coupling matrix $\Hab$,   electron transmission coefficient $\kel$, transfer probability $\PLZ$, reorganization energy $\lambda$ and hopping barrier $E_a$.}
        \label{table:hopping-parameters}
    
        \begin{tabular}{ccc}
            \hline
            Hopping & 1NN & 2NN \\
            \hline
            Distance \AA & 3.9 & 5.0 \\
            $\Hab$ (meV) & 91 & 24 \\
            $h\veff$ (meV) & 276 & 297 \\
            $\lambda$ (eV) & 1.20 & 1.27 \\
            $\PLZ$ & 0.60 & 0.057 \\
            $\kel$ & 0.75 & 0.11 \\
            $E_a$ (meV) (NEB) & 250 & 269 \\
            $k_{\textrm{ET}}$ ($s^{-1}$) & $4\times10^9$ & $3\times10^8$ \\
            \hline
        \end{tabular}
    \end{table}

    \subsubsection{Small polaron mobility $\mu$ for pristine \ce{BiVO4}}
    
    
     Computing the polaron hopping mobility from kMC simulations can easily take into account the anisotropicity and 2NN hopping, instead of using an analytic formula where only one barrier can be included as with most of the past work\cite{Rettie2015,Rettie2016,Kweon2015,Wang2016}. We always included the 1NN hopping that has 3.9 {\AA} distance between two hopping centers and has the smallest barrier. Meanwhile, we also considered the 2NN hopping which has  5 {\AA} distance and a comparable hopping barrier to 1NN as shown in Table \ref{table:hopping-parameters}.
    Interestingly, from Landau-Zener theory we found the 2NN hopping charge transfer rate $k_{ET}$ is less than 1/10 of 1NN in Table \ref{table:hopping-parameters},  so the 2NN hopping has an insignificant effect on the mobility by kMC simulations as shown in Table \ref{table:compare-mobility-direction}. Therefore, 2NN hopping can be safely neglected in the mobility simulation of BiVO$_4$. With the computational techniques and numerical inputs discussed above, we obtained the mobility of pristine \ce{BiVO4} in reasonably good agreement with the experimental results of lightly doped \ce{BiVO4} shown in Table \ref{table:mobility}. Previous studies with kinetic Monte Carlo simulation\cite{Liu2018} significantly overestimated the barrier with the linear interpolation method and thus
    likely underestimated the carrier mobility. In addition, the polaron transport process was assumed to be fully adiabatic in previous studies of \ce{BiVO4}\cite{Rettie2015}, where $\kel$ is approximated as 1. 
    We note that this assumption is not reliable in \ce{BiVO4}, which could lead to qualitatively wrong results, such as the mobility ratio along a and c lattice directions $\mu_a/\mu_c$ as discussed below.
    \begin{table}
    \begin{threeparttable}
        \caption{Electron drift mobility of pristine and doped \ce{BiVO4} from experiments and first-principle calculations at room temperature 300 K.}
        \label{table:mobility}
        \begin{tabular}{ccc}
          \hline
          System & Method & \vtop{\hbox{Electron Drift Mobility} \hbox{($\textrm{cm}^2\textrm{V}^{-1}\textrm{s}^{-1}$)}} \\
          \hline
          0.3\% W\tnote{1} & Experiment\cite{Rettie2015} & $5\times10^{-5}$ \\
          1\% W\tnote{2} & Experiment\cite{Abdi2013} & $2.2\times10^{-4}$ \\
          Pristine & This work & $1.38\times 10^{-4}$ \\
          3\%Mo(W) & This work & $1.07\times 10^{-4}$ \\
          6\%Mo(W) & This work & $0.91\times 10^{-4}$ \\
          \hline
        \end{tabular}
        \begin{tablenotes}
          \item[1] Deduced from DC conductivity and Seebeck coefficient
          \item[2] Measured combined electron and hole mobility from time-resolved microwave conductivity
        \end{tablenotes}    
    \end{threeparttable}
    \end{table}
    
    It has been experimentally observed that hopping conductivity of monoclinic \ce{BiVO4} is anisotropic\cite{Rettie2015}. Anisotropicity of carrier conduction has been found in other metal oxides as well, mainly due to specific geometric characteristics such as a layered structure\cite{Adelstein2014}. The anisotropicity is also observed in our kMC simulation as shown in Table \ref{table:compare-mobility-direction}.
    However, in \ce{BiVO4}, there is no such prominent geometry feature, thus this anisotropic mobility must be related to more subtle structural differences  
    among three lattice directions in \ce{BiVO4}.
    Based on a simple geometric relation (details can be found in SI), when only the nearest neighbor hopping is considered, the square of displacement $L^2$ along $a$- or $b$- axis on average is only 0.38 times of that along $c$-axis in \ce{BiVO4}.  Since the diffusion coefficient $D$ is proportional to $L^2$ (Eq. \ref{eq:diffusion-rw}),  $D$ or the mobility $\mu$ (linearly proportional to $D$ in Eq.~\ref{eqES}) along $a$- or $b$- axis is only 0.38 of $c$- axis, which agrees with our kMC simulation in Table \ref{table:compare-mobility-direction}.
    
    The 2NN hopping is in the $ab$-plane, so the faster the 2NN hopping is, the larger $\mu_a/\mu_c$ mobility ratio will be. If both 1NN and 2NN hoppings are assumed to be fully adiabatic with $\kel\approx 1$, the kMC simulation will give $\mu_a/\mu_c=1.13$, which is qualitatively wrong. This is because the 2NN hopping has a $\kel$ 7 times smaller than 1NN, which will give a smaller charge transfer rate (see Table \ref{table:hopping-parameters}) and a small contribution to carrier mobility. With correct $\kel$ for both 1NN and 2NN hopping rates, the carrier mobility did not change much after we added 2NN hopping; therefore we will neglect 2NN hopping in the next sections. 
    
    \begin{table}
        \caption{Drift mobility along different axes with and without second nearest-neighbor hopping in the $ab$-plane at 300K. 1NN denotes the first nearest neighbor hopping and 2NN denotes the second nearest neighbor hopping. Note that with $\kel=1$, the $\mu_a/\mu_c$ ratio including 2NN (1NN+2NN) is significantly overestimated compared with full Landau-Zener theory (with computed $\kel$).}
        \label{table:compare-mobility-direction}
        \begin{tabular}{cccccc}
          \hline
          Neighbor  & $\kel$  & \multicolumn{4}{c}{Mobility ($10^{-4}\textrm{cm}^2$/V/s)}\\
            & & Avg. & $ab$-plane & $c$-axis & $\mu_a/\mu_c$ \\
          \hline
          Only 1NN & 0.75 & 1.38 & 0.90 & 2.35 & 0.38 \\
          1NN+2NN  & 0.75/0.11 & 1.55 & 1.15 & 2.34 & 0.49\\
          1NN+2NN & 1/1 & 3.06 & 3.19 & 2.82 & 1.13 \\
          \hline
        \end{tabular}
    \end{table}
    
    \subsection{Small Polaron Mobility of Doped \ce{BiVO4}}
    \subsubsection{Polaron energies as a function of dopant-polaron distances}
    \begin{figure}
        \centering
        \includegraphics[width=0.8\linewidth]{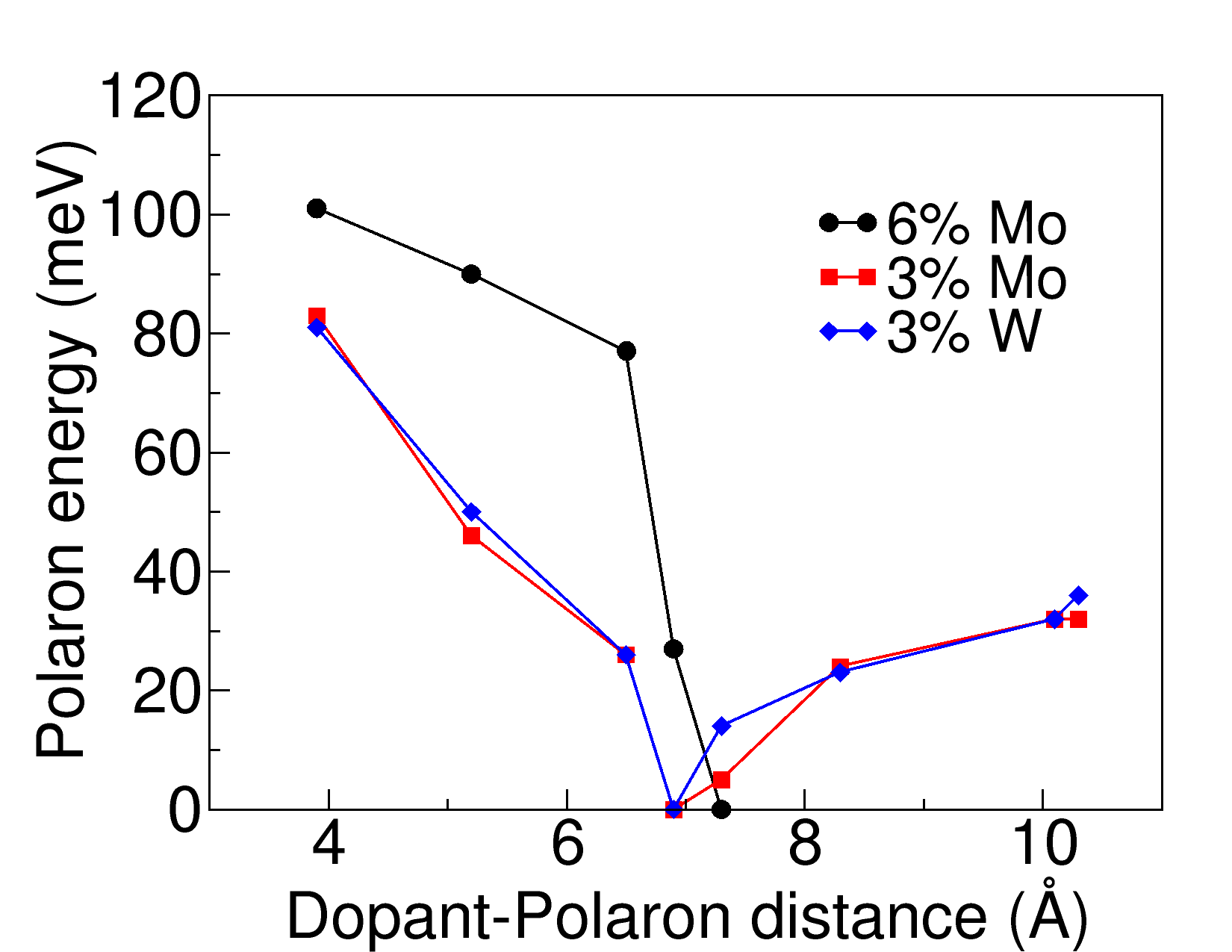}
        \caption{Total energies of polaronic states as a function of dopant-polaron distances in Mo (3$\%$, 6$\%$) and W (3$\%$) doped supercells. 3$\%$ doping corresponds to one dopant per 192 atom supercell (32 \ce{BiVO4} units).
        All values are referenced to the most stable site with dopant-polaron distance around 7 {\AA} (with the lowest total energies).}
        \label{fig:polaron-ground-state-energy}
    \end{figure}
    
    In this section we will discuss the polaron-dopant interaction and understand its effect on the mobility and underlying mechanism, which is critical to further design of materials with improved carrier mobility. 
    Here we chose three n-type representative dopants Cr, Mo and W substitution of V atoms as examples to compare their effects on polaron hopping transport properties. The structural models are constructed on the basis of the chemical formula \ce{BiV_{1-x}M_xO4}, where $x$ is the dopant concentration and M can be Cr, Mo or W. We chose 3\% and 6\% doping concentrations as two examples in order to study the effect of doping concentration on polaron transport. The models are a 96-atom (6$\%$) or a 192-atom (3$\%$) supercell with a V atom substituted by a dopant atom. We note that at 6$\%$, we expect significant dopant-dopant interaction, different from a dilute limit.
    
    To understand the nature of polaron-dopant interaction, we first compared the total energies/stability when the extra electron from n-type dopants localizes at different V or dopant sites. 
    In particular, for the case of Cr, we found that the excess electron from Cr can only be stabilized at the Cr atom and form a filled gap state that is mainly composed of Cr $d$ orbitals just below the conduction band 
    (as shown in Projected Density of State and gap state wavefunction in SI. In other words, 
    the electron from Cr cannot be ionized easily and form a stable electron polaron at V, similar to the findings in Ref.\citenum{Park2011}. This is due to the highly localized 3d orbitals of Cr atom. Therefore, Cr has an oxidation state of 5+ and is a donor (that potentially forms 6+ state) with a very high ionization energy as discussed in Ref.\citenum{Yin2011}.  As a result, Cr acts as an electron trap and electron-hole recombination center.
    
    In contrast, for the case of Mo and W doping, one electron is spontaneously ionized from Mo/W, localizes at the V site and forms a small polaron accompanied by local lattice distortions. In other words, the extra electron from Mo/W (as $n$-type dopants) is thermodynamically stable to localize around a V site to form small polarons instead of the dopant sites.
    The interaction between Mo/W dopants and electron polaron can be understood 
    from the change of total energies as a function of distances between the dopant and polaron in Fig. \ref{fig:polaron-ground-state-energy}.  
    We can identify two shells of neighbor sites around one \ce{MoO4} or \ce{WO4} tetrahedral with different trends. In the first shell (with Mo/W-V distances between 3-7 {\AA}), the total energy decreases
    as a function of the dopant-polaron distances; therefore, the interaction between the two is repulsive and the polaron prefers to move away from the Mo/W dopant. 
    In the second shell (with Mo/W-V distances between 7-11 {\AA}), the total energy increases slightly as a function of dopant-polaron distances which indicates a weak attractive interaction. 
    Outside the second shell, the interaction between a polaron and a dopant is negligible, so the formation energy recovers the bulk limit. The boundary between the two shells is approximately 7 {\AA} (where the minimum total energy in Fig.~\ref{fig:polaron-ground-state-energy} is used as the reference zero), which is already the largest dopant-polaron distance in 6$\%$ Mo doping supercell so the second shell exists only in lower concentration systems, \textit{e.g.} the 3$\%$ doping case. We note that the total energies as a function of dopant-polaron distances in 3$\%$ W doped \ce{BiVO4} have very similar values to the case of 3$\%$ Mo doping (reference to the polaron energy minimum at 7 {\AA}) as shown in Fig.~\ref{fig:polaron-ground-state-energy}. 
    
    In general, the ionized n-type dopants (which are positively charged) and electron polarons have attractive electrostatic interactions, which should not facilitate the polaron conduction in the crystal.
    The effect that counters the electrostatic attraction stems from the local lattice distortion of dopants and polarons. Specifically, in pristine and doped \ce{BiVO4}, when a polaron formed at a \ce{VO4} site, the V-O bond length is stretched by 0.1 {\AA}. Meanwhile, the Mo-O or W-O bond length (even after being ionized) is  0.06 {\AA} longer than the V-O one without a polaron. Two larger tetrahedra are energetically unfavorable to stay close, in order to minimize the local lattice distortions. We would expect this effect to decrease faster than electrostatic interactions as the bond energy scales as $\approx r^2$ (where $r$ is the bond length) near equilibrium positions in the harmonic approximation. On the other hand, the electrostatic attraction being a long-range interaction decreases as $r^{-1}$. 
    As a combination of two counteracting effects, the lattice distortion dominates at a short polaron-dopant distance and electrostatic attraction dominates at a long polaron-dopant distance, which correspond to the two shells we showed in Fig. \ref{fig:polaron-ground-state-energy} respectively; then the energy minimum appears at the boundary between the first and the second shell. 
    The energy required to move a polaron from the energy minimum (7 {\AA} to the dopant) to the bulk region is only approximately 30 meV, which indicates polarons can move away from this energy minimum easily at room temperature. Overall, though all three dopants (Cr, Mo and W) are n-type, the interaction between Mo/W and polarons is dominated by a "repulsive" interaction, which is opposite to Cr being an electron "trap"; the different types of interaction determine whether dopants will facilitate or hinder the polaron transport.
    
    \subsubsection{Polaron mobility of doped \ce{BiVO4} with kMC samplings}
    
    Through coupling charge transfer rates by Landau-Zener theory (Eq.~\ref{eq:lz-rate}) and kMC sampling, we for the first time simulated the polaron mobility in the presence of dopants under this framework fully from first-principles.
    We obtained optimized polaron structures at all non-equivalent sites and computed hopping rates between all first nearest neighbor (1NN) pairs, and then used them as inputs for kMC simulations of hopping mobility.
    As discussed earlier, only 1NN is necessary for mobility calculations and 2NN has negligible contributions,  therefore,
    all the 1NN hopping barriers were computed by NEB at the DFT+U level, and $\veff$ and $\PLZ$ were kept at the same values as the pristine systems.
     
\begin{figure}
        \centering
        \includegraphics[width=0.9\linewidth]{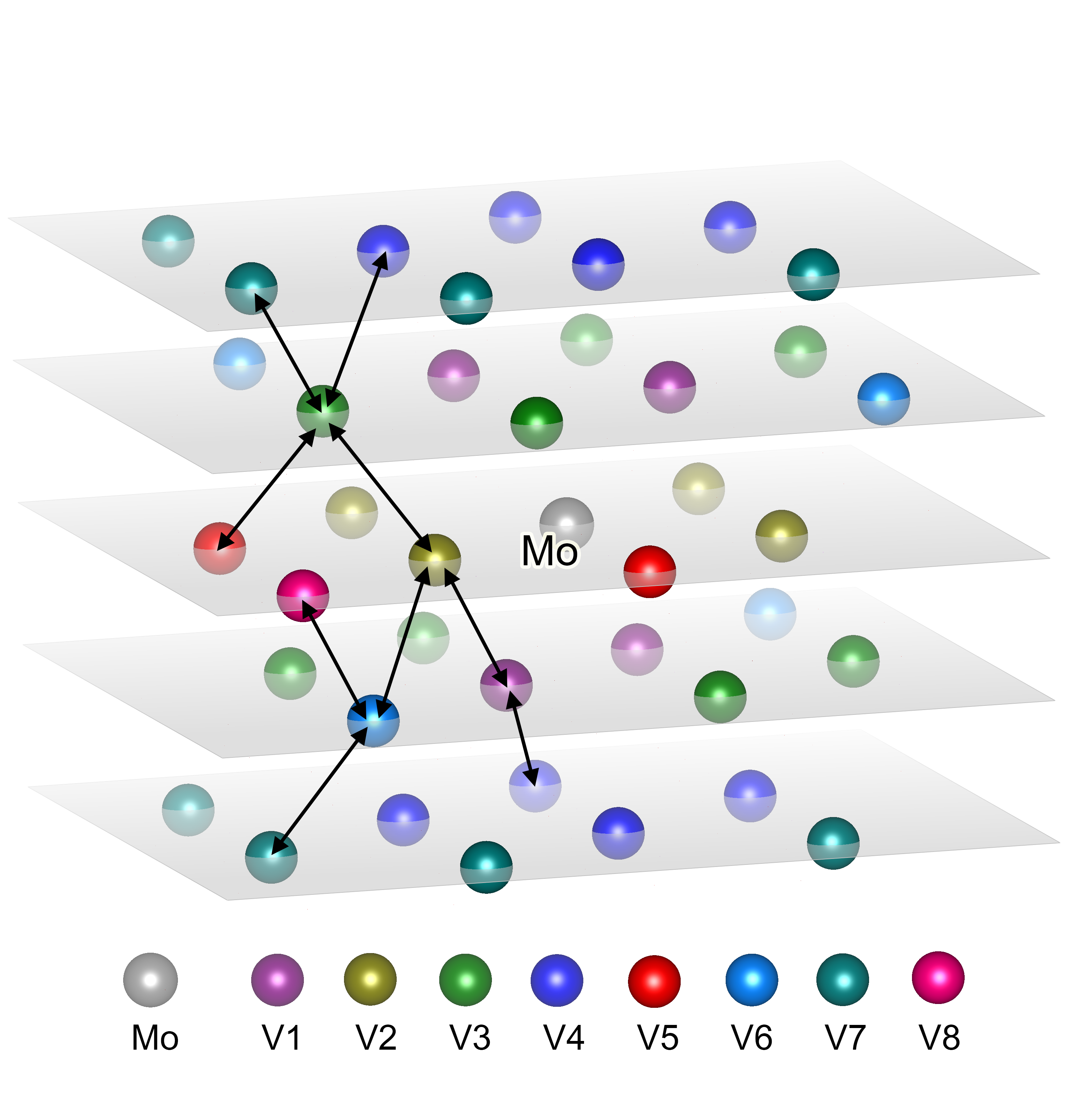}
        \caption{V atoms in 3\% Mo doped \ce{BiVO4} supercell. Hopping paths listed in Table \ref{table:32-neb-barrier} are marked as arrows. V1 to V8 are sorted in an ascending order based on their distances to the nearest Mo atom (considering the periodic boundary condition). Equivalent V atoms are marked with the same color. For simplicity, only one of equivalent hopping paths is shown in the figure. 
        }
        \label{fig:polaron-distance-dopant-arrow}
    \end{figure}

    Multiple nonequivalent hopping paths exist in the doped system (3$\%$ Mo doping) when the periodic boundary condition is applied, as shown in Fig. \ref{fig:polaron-distance-dopant-arrow}. The corresponding barriers obtained by the NEB method are listed in Table \ref{table:32-neb-barrier}. They are no longer symmetric as pristine \ce{BiVO4}; instead, generally 
    along one hopping direction (\textit{e.g.} left side$(L)\rightarrow$ right side$(R)$) the barrier is lower than the one in pristine, and along the reversed hopping direction $R\rightarrow L$, the barrier is higher. This is because the interaction between the Mo/W dopants and small polarons is repulsive at a short range as discussed in the previous section, which leads to a lower barrier to hop away from the dopant and a higher barrier to hop towards the dopant. We also found the barriers between two directions ($L\rightarrow R$ and $R\rightarrow L$) in Table \ref{table:32-neb-barrier} become closer when the distance between a dopant and a small polaron is larger, due to a weaker dopant-polaron interaction. Eventually a value close to the pristine bulk hopping barrier will be recovered when dopants and polarons are far enough from each other.
    
    \begin{table}
        \caption{Hopping barriers computed by NEB along two directions at different sites of 3\% Mo doped BiVO$_4$. $L$ and $R$ refer to the left and right sides of $L\leftrightarrow R$ in the first column. Site names V$i$ are shown in Fig. \ref{fig:polaron-distance-dopant-arrow} which are defined by the distances to the Mo dopant.}
        \label{table:32-neb-barrier}
        \begin{tabular}{ccc}
          \hline
          Sites($L\leftrightarrow R$) & $E_a(L\rightarrow R)$ (meV) & $E_a(R\rightarrow L)$ (meV) \\
          \hline
          Pristine & 250  & 250  \\
          V1$\leftrightarrow$V2 & 231  & 268  \\
          V1$\leftrightarrow$V4 & 213 & 296  \\
          V2$\leftrightarrow$V3 & 236  & 255  \\
          V2$\leftrightarrow$V6 & 240  & 260  \\
          V3$\leftrightarrow$V4 & 243  & 268  \\
          V3$\leftrightarrow$V5 & 240  & 260  \\
          V3$\leftrightarrow$V7 & 250  & 242  \\
          V8$\leftrightarrow$V6 & 252  & 243  \\
          V7$\leftrightarrow$V6 & 254  & 241  \\
          \hline
        \end{tabular}
    \end{table}
    
    Due to the broken symmetry in the presence of dopants, the carrier mobility of doped systems requires statistical samplings of hopping rates along all possible pathways with periodic boundary conditions. The kinetic Monte Carlo simulation with the barriers in Table~\ref{table:32-neb-barrier} as inputs is performed to obtain the electron mobility in pristine and doped system. The details of the kMC samplings can be found in SI. An effective barrier can be defined from mobilities at different temperatures as $\mu(T)=A\exp(-E_{\textrm{eff}}/k_BT)$. At room temperature, the effective barrier is 250 meV for pristine system and 267 meV for 3\% Mo doped system with part of hopping paths shown in Fig. \ref{fig:polaron-distance-dopant-arrow}. The computed mobilities are listed in Table \ref{table:mobility}. 
    Our computed carrier mobility has reasonably good agreement with experimental results\cite{Abdi2013,Rettie2015}, which validate our methodology and numerical implementation. 
    Overall the Mo or W doping (3 $\%$) did not affect the mobility significantly from our calculations (slightly decreased from the pristine systems), for which the underlying physics will be discussed in detail below.
    
    \begin{figure}
        \centering
        \includegraphics[width=1.0\linewidth]{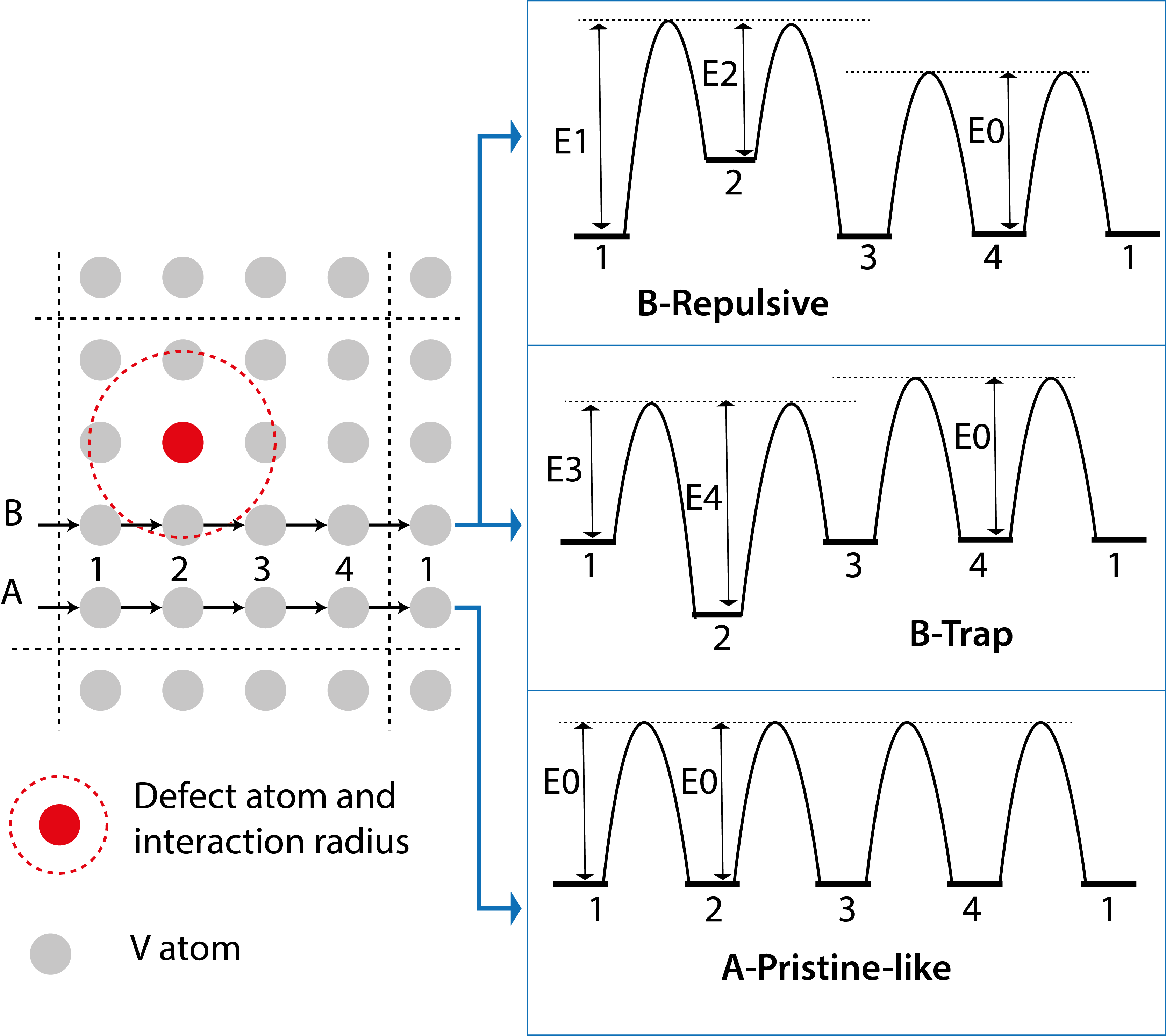}
        \caption{Schematic diagram of polaron transport processes in doped \ce{BiVO4} with periodic boundary conditions. The red dashed circle shows the region where the dopant-polaron interaction is non-negligible. Polarons hopping along Pathway A are not affected by dopants while the ones along pathway B are affected. The polaron-dopant interaction can be repulsive or attractive (trap), so there are three kinds of pathways in total: (A-Pristine-like), (B-Repulsive) and (B-Trap). $E_1,E_3$ are the barriers to jump into the interaction region, $E_2, E_4$ are the barriers to jump out of this region and $E_0$ is the hopping barrier in the pristine system. For the dopant-polaron repulsive interaction (\textit{e.g.} Mo and W doping) we have $E_1>E_0>E_2$ and for the attractive interaction (\textit{e.g.} Cr doping as a trap) we have $E_3 < E_0 < E_4$. Therefore, to pass this interaction region, a polaron must overcome a larger barrier and a smaller barrier than $E_0$.}
        \label{fig:defect-repulsive-trap}
    \end{figure}
    
    In general, polaron transport pathways in a doped system can be classified in two groups as shown in Fig.~\ref{fig:defect-repulsive-trap}: (A) polarons which do not cross regions that have interaction with dopants (represented by a red dashed circle in Fig.~\ref{fig:defect-repulsive-trap}) and (B) polarons which pass through those interaction regions. For (A), all hopping barriers along the pathway are close to the pristine system so the overall transport rate also recovers the pristine limit, which is referred to as ``A-Pristine-like" in Fig.~\ref{fig:defect-repulsive-trap}.
    
    For the group (B), when the dopant-polaron interaction is attractive (\textit{i.e.} along the ``B-Trap" pathway in Fig.~\ref{fig:defect-repulsive-trap}), the polaron will move closer to dopants with a lower barrier (E3 in Fig.~\ref{fig:defect-repulsive-trap})  compared with the barrier in pristine systems. The first step determines if the polaron will prefer to move along pathway ``B-Trap" instead of ``A-Pristine-like" due to a low barrier. Then the second step with a higher barrier than pristine (E4 in Fig.~\ref{fig:defect-repulsive-trap}) is the rate-determining step and causes the hopping rate along this pathway ``B-Trap" to be slower than ``A-Pristine-like", or polarons could not even get out of the trap position 2 at room temperature. 
    So the overall mobility along ``B-Trap" will be lower than pristine system, and the dopants act as a "trap" of the polaron, such as the case of Cr doping. 
    
    When the dopant-polaron interaction is repulsive (\textit{i.e.} along the ``B-Repulsive" pathway in Fig.~\ref{fig:defect-repulsive-trap}), the polaron must overcome a higher barrier (E1) to move closer to the dopant and then move further with a lower barrier (E2) compared with the pristine barrier. This high barrier step (E1) slows down the overall hopping rate of this pathway, and also lowers the probability of choosing this pathway ``B-Repulsive". As a result, if the ``A-Pristine-like" pathway exists in the sample, it will dominate the transport process, which means the mobility will recover that of the pristine system. This is the case for Mo-doped \ce{BiVO4} in our kMC simulation .
    
    Therefore for such simulations with only one polaron and one dopant in a supercell, once all hopping paths in an infinitely large system are considered, one can either get a smaller mobility when the dopant is a trap (``B-Trap" in Fig.~\ref{fig:defect-repulsive-trap})  or a mobility similar to the pristine bulk when the dopant has a repulsive interaction with the polaron (``B-Repulsive" in Fig.~\ref{fig:defect-repulsive-trap}), if the bulk region is recovered in the supercell, \textit{i.e.} ``A-Pristine-like" path exists. 
    
    This conclusion holds only at lower doping concentration than 6\% Mo or W, where the regions affected by dopants (``interaction radius" in Fig.\ref{fig:defect-repulsive-trap} is around 7-9 {\AA} based on calculations in Fig.~\ref{fig:polaron-ground-state-energy}) do not overlap with each other, assuming
    the dopants are homogeneously distributed in the material. 
    This will allow for ``A-Pristine-like" pathway as there is no pristine-like sites in 6\% supercell based on our calculations in Fig.~\ref{fig:polaron-ground-state-energy}. In experiments the dopants are not necessarily evenly distributed, where ``A-Pristine-like" pathways may be possible even at a higher concentration than 6\%. \cite{Abdi2013,Rettie2013} The above discussions described the physical pictures and explained underlying mechanism of our computed results in Table~\ref{table:mobility}.
    
\begin{figure}
        \centering
           \includegraphics[width=1.0\linewidth]{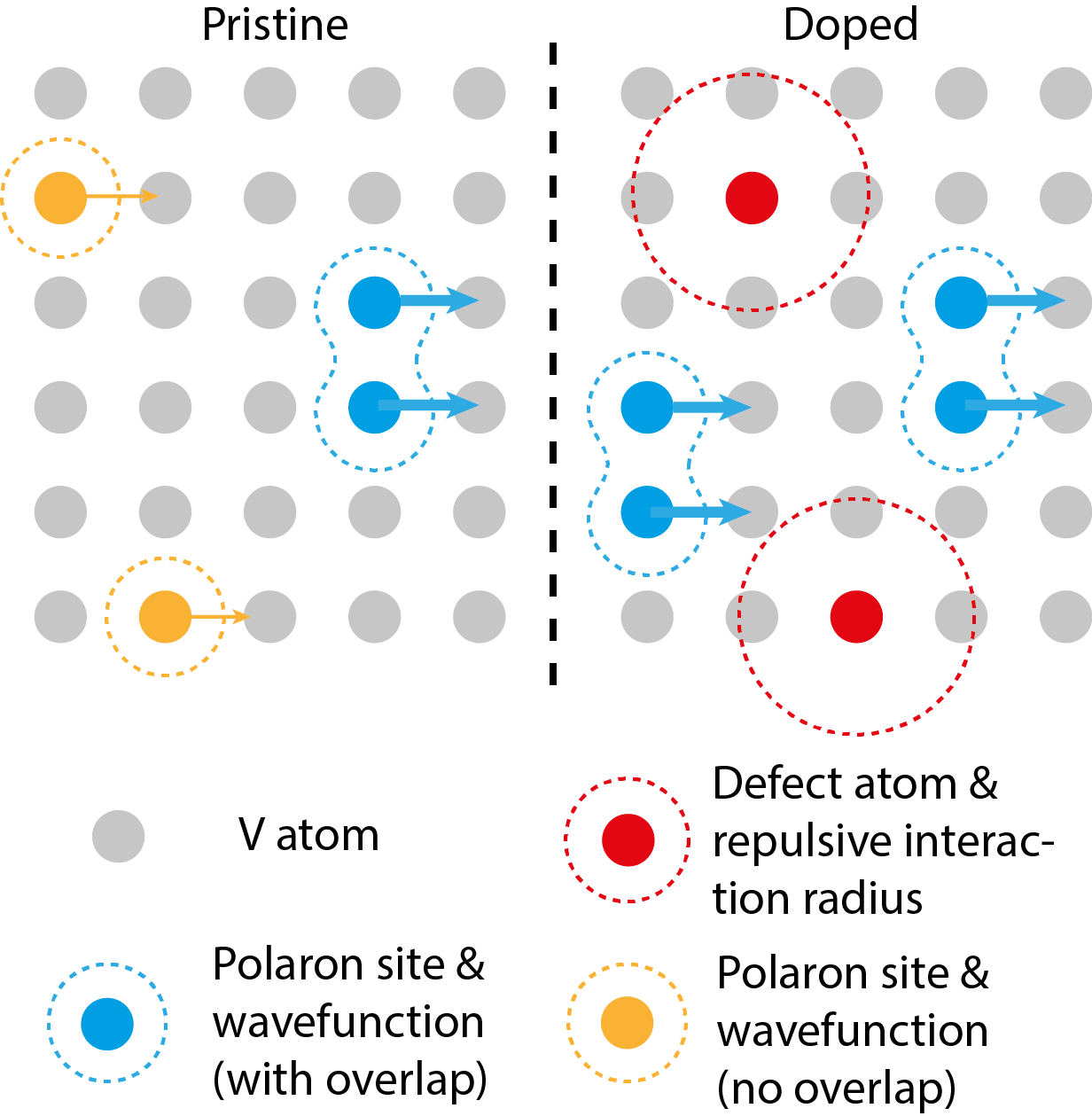}
        \caption{Schematic diagram showing how dopants having a repulsive interaction with polarons can boost the polaron transport through locally concentrated polarons. We considered the same number of polarons in the pristine system (left) and the doped system with repulsive dopant-polaron interactions (right). Because dopants like Mo/W push away polarons to regions distant to all dopants, polarons have a higher local concentration at such regions. Therefore polarons can have larger wavefunction overlaps that may form more delocalized wavefunctions (shown as a larger blue region in the figure) that may improve the hopping conduction.}
        \label{fig:defect-concentration-effect}
    \end{figure}

    In addition, the polarons may not be homogeneously distributed, even if the dopants are evenly distributed; instead, in the presence of dopants similar to the case of W and Mo doping (\textit{i.e.} the polaron-dopant interaction is repulsive), polarons are likely  pushed away by dopants and concentrated in regions distant from most dopants,  as shown in  
    Fig. \ref{fig:defect-concentration-effect}.
    This effect may play an important role in the hopping mobility but has not been included in our supercell calculations: polaron wavefunctions may overlap and become more delocalized which can lower the polaron hopping barriers\cite{Kweon2015}. At the highly concentrated polaron limit, the band conduction with completely delocalized electrons may be recovered. 
    Therefore, the carrier mobility we obtained for Mo and W doped samples represents the low limit (in the absence of other dopants or defects in the samples), which can be higher in experiments due to inhomogenous polaron and dopant distributions. 
    As discussed in the introduction, experimentally whether the electron mobility increased or decreased in the presence of Mo/W dopants is still controversial. Our results may explain the physical reason for this controversy: depending on the doping concentration and distributions, one may get lower, similar or higher hopping mobility compared to the pristine systems. Another complication is that the oxygen vacancy may also affect the hopping mobility significantly (whose concentration may not be the same at pristine and doped systems). But it is difficult to quantify its concentration experimentally, which could also lead to inconsistency between different experimental results.

    \section{Conclusion and Outlook}
    
    In conclusion, we established the theoretical framework of coupling the Landau-Zener theory and kinetic Monte Carlo (kMC) simulations to compute hopping mobility for anisotropic and doped systems fully from first-principles. We used \ce{BiVO4} as an example where we obtained electron mobility in good agreement with experimental measurements. We showed that the statistical samplings of hopping trajectories are critical for anisotropic systems and especially important for doped systems, where the symmetry of the bulk systems is broken.

    The electron polaron transport in \ce{BiVO4} is neither fully adiabatic nor nonadiabatic, and the correct description of the polaron hopping rate and anisotropicity demands the Landau-Zener theory instead of classical transition state theory or the Marcus theory in the corresponding adiabatic and nonadiabatic limit. From the Landau-Zener theory, the 1NN hopping has a much larger hopping rate than the 2NN one due to much smaller
    electronic couplings and $\kel$ in the latter case, although their hopping barriers are comparable. Without taking into account $\kel$ explicitly in the rates and assuming adiabatic transfer for both 1NN and 2NN hoppings will result in qualitatively wrong mobility.
    In addition, the electron mobility in pristine BiVO$_4$ shows strong anistropicity, which requires statistical samplings like kMC instead of an analytical formula with one effective barrier.
    
    With this approach, we also studied the doping effect on the polaron transport properties at the microscopic level, by using Cr, Mo, W doped BiVO$_4$ as examples.
    We showed that in the case of \ce{BiVO4}, the Mo/W dopant acts as a "repulsive" center and polarons will be pushed away from the dopant outside the dopant-polaron repulsive region with a radius around 7 {\AA}. 
    This is because both Mo/W substitution of V atoms and electron polaron formation locally expand the lattice, which
    create a short-ranged repulsive interaction between the two in order to minimize the local strain, despite the long-range Coulomb attraction between an ionized Mo/W dopant (positively charged) and an electron polaron (negatively charged).
    On the other hand, Cr acts as a strong trap of electrons and will lower the hopping mobility and conductivity.
    The nature of dopant-polaron interactions such as a repulsive interaction, characterized by total energy changes as a function of polaron-dopant distances can be used as an important descriptor to screen the promising dopants that can potentially overcome low hopping mobility in polaronic oxides.
    
    For polaron mobility calculations of doped materials, we found a mobility  either less or equal to that in pristine systems will be obtained,   as long as the dopant and polaron concentration is relatively low and homogeneously distributed, \textit{i.e.} numerically,
    one dopant and one polaron are considered in the simulated supercell with periodic boundary conditions. 
    This represents a lower bound of the hopping mobility, considering polarons may be concentrated in small regions distant from all dopants if dopants and polarons have repulsive interactions. 
    The overlap of polaron wavefunctions and formation of delocalized states can lower the hopping barriers, improve the hopping mobility and even change the nature of conduction.
    
    Therefore, to boost small polaron conduction in polaronic oxides, "good dopants" should be able to increase the overall electronic conductivity  following the criteria below: a) being a shallow dopant with low ionization energies such as W/Mo in BiVO$_4$, which can increase carrier concentration at room temperature; b) having a "repulsive" interaction with the polarons instead of an attractive interaction, which can easily hop away from the dopants, and in that case the computed mobility should be similar to the pristine systems at the homogeneous distribution of dopants and polarons.
    
    Future work requires simulations with multiple dopants and polarons in a supercell and compute dynamical electronic couplings and hopping rates 
    depending on polaron-polaron distances (taking into account polaron wavefunction overlaps quantum mechanically),
    which can provide a further understanding of the effect of inhomogeneous distribution of dopants/polarons on polaron transport in both pristine and doped materials. We note that our framework by coupling the Landau-Zener theory and kMC is an important forward step towards simulating hopping mobility in anisotropic and doped systems from first-principles, and understand the doping effect on polaron mobility at the microscopic level. 
    
    \section*{Acknowledgement}
    We thank very useful discussions with Alexei Stuchebrukhov and Giulia Galli. Y.P. acknowledges the financial support from the National Science Foundation under grant No. DMR-1760260 and Hellman Fellowship. This research used resources of the Center for Functional Nanomaterials, which is a U.S. DOE Office of Science Facility, and the Scientific Data and Computing Center, a component of the Computational Science Initiative, at Brookhaven National Laboratory under Contract No. DE-SC0012704.
    This research also used resources of
    the National Energy Research Scientific Computing Center (NERSC), a DOE Office of Science User Facility supported by the Office of Science of the US Department of Energy under Contract No. DEAC02-05CH11231, the Extreme Science and Engineering Discovery Environment (XSEDE), which is supported by National Science Foundation under grant No. ACI-1548562. \cite{xsede}
    
    
    \bibliography{main,main2,package} 

\end{document}